\documentclass[pdflatex,sn-mathphys-num]{sn-jnl}

\usepackage{graphicx}%
\usepackage{multirow}%
\usepackage{amsmath,amssymb,amsfonts}%
\usepackage{amsthm}%
\usepackage{mathrsfs}%
\usepackage[title]{appendix}%

\usepackage{textcomp}%
\usepackage{manyfoot}%
\usepackage{booktabs}%
\usepackage{algorithm}%
\usepackage{algorithmicx}%
\usepackage{algpseudocode}%
\usepackage{listings}%
\usepackage{float} 

\usepackage{placeins}

\usepackage{adjustbox}
\usepackage{eqparbox}
\usepackage{array}
\usepackage{soul}
\usepackage{icomma}
\usepackage{marvosym}
\usepackage{textcomp}

\usepackage[table,xcdraw]{xcolor} 

\usepackage{tabularx} 

\usepackage{caption}
\usepackage{subcaption}
\usepackage{float}

\theoremstyle{thmstyleone}%
\theoremstyle{thmstyletwo}%

\theoremstyle{thmstylethree}%

\begin{document}

\title{Eigenvector-Based Sensitivity Analysis of Contact Patterns in Epidemic Modeling.}


\author*[1,2]{\fnm{Evans Kiptoo} \sur{Korir}}\email{evanskorir6@gmail.com}

\author[1,2]{\fnm{Zsolt} \sur{Vizi}}\email{
zsvizi@math.u-szeged.hu}
\equalcont{These authors contributed equally to this work.}

\affil*[1]{\orgdiv{Bolyai institute}, \orgname{University of Szeged}, \city{Szeged}, \postcode{6720}, \country{Hungary}}

\affil[2]{\orgname{National Laboratory for Health Security}, \state{Szeged}, \country{Hungary}}


\abstract{
Identifying which age groups contribute most to uncertainty in disease transmission models is essential for improving model accuracy and guiding effective interventions. This study introduces an eigenvector-based sensitivity analysis framework that quantifies the influence of age-specific contact patterns on epidemic outcomes. By applying perturbation analysis to the Next Generation Matrix, we reformulate the basic reproduction number, \(\mathcal{R}_0\), as an eigenvalue problem, allowing us to pinpoint the age group interactions most critical to transmission dynamics. While the framework is broadly applicable to clinical outcomes such as hospitalizations, ICU admissions, and mortality, we focus here exclusively on the mortality outcome. We demonstrate the approach using two age-structured COVID-19 models with contact matrices from the UK and Hungary. Our results illustrate how differences in demographics, contact structures, and age group aggregations shape model sensitivity.
}




\maketitle

\section{Introduction}\label{sec1}

Understanding how infectious diseases spread through populations requires an accurate characterization of social contact patterns, which form the foundation of transmission in compartmental epidemic models \cite{sensitivity}. 
The \textit{social contact hypothesis} \cite{wallinga} states that the number of secondary infections generated by an individual is proportional to their social contacts, with a pathogen-specific proportionality factor.
Age-structured contact matrices, estimated by numerous studies~\cite{adu2022quantifying, ajelli2017estimating, grijalva2015household, horby2011social, kiti2014quantifying, read2014social, le2018characteristics, melegaro2017social, mossong2008, prem}, are widely used to represent the average number of interactions between individuals in different age groups, reflecting the heterogeneity of real-world social mixing~\cite{mossong2008, prem}. These matrices are essential for capturing how interventions, susceptibility, and behavioral factors influence disease dynamics.  
However, contact patterns are inherently uncertain and vary across time, geographic regions, and population demographics. During the COVID-19 pandemic, for instance, behavioral changes and non-pharmaceutical interventions (NPIs) such as mobility restrictions, lockdowns, and school closures caused substantial and often uneven shifts in contact rates~\cite{bokanyi, munday}. Furthermore, many contact matrices are derived from limited survey samples or extrapolated data~\cite{fumanelli}, introducing additional uncertainty. These challenges can significantly impact model predictions and, consequently, public health decision-making.

Given these limitations, it is crucial not only to incorporate contact data into epidemic models but also to systematically assess how sensitive key epidemiological outcomes, such as the basic reproduction number (\(\mathcal{R}_0\)), hospitalizations, and mortality, are to variations in the underlying contact structure. Sensitivity analysis provides a structured way to do this. By identifying the contact matrix elements that most strongly influence model outcomes, such analysis enables researchers to prioritize data collection efforts and refine public health interventions \cite{sensitivity}.

Traditional global sensitivity techniques, such as Monte Carlo simulations, Latin Hypercube Sampling (LHS), and Partial Rank Correlation Coefficients (PRCC)~\cite{hamby1994review, helton2000sampling, morris1992factorial, sobol1990sensitivity, sensitivity}, can capture uncertainty across parameter spaces but are computationally expensive and less informative about matrix-structured inputs like contact patterns. In contrast, analytical methods based on the Next Generation Matrix (NGM) provide more targeted insight into how changes in contact rates propagate through a model. Eigenvalue-based sensitivity analysis, in particular, allows one to assess the influence of contact perturbations on \(\mathcal{R}_0\) by differentiating the NGM with respect to the contact matrix. For instance, Angeli et al.~\cite{angeli2024acquires} applied formal perturbation techniques to the NGM to analyze the early spread of COVID-19 in Belgium. These approaches are especially useful in models where parameter interactions are more linear and tractable analytically, though they may be less effective in highly nonlinear or stochastic frameworks.

In this study, we extend the analytical approach proposed in~\cite{angeli2024acquires} by developing a general sensitivity analysis framework centered on age-structured contact matrices. Unlike prior work, we construct a computational graph with the independent entries of the contact matrix as inputs, allowing us to explicitly trace how these elements influence key outcomes. Using the dominant eigenvalue of the NGM as a proxy for transmission potential, we can compute the gradient of \(\mathcal{R}_0\) with respect to each contact element. We extend this framework beyond transmission analysis by linking the NGM to downstream clinical outcomes such as hospitalization, ICU admission, and mortality. Additionally, we demonstrate the flexibility of our framework by applying it to other epidemic models and contact data from the literature, showcasing its utility in various settings.

\section{Methods}\label{sec11}

\subsection{Framework for Sensitivity Analysis}

The sensitivity analysis framework is designed to flexibly assess the impact of changes in contact patterns on disease transmission across various structured epidemic models. By updating key input files, users can adapt the framework to accommodate different social contact matrices, varying levels of model complexity, and customized model parameter values.

The framework’s primary input is a set of full and aggregated contact matrices that represent age-specific social interactions across various settings (e.g., homes, schools, workplaces, and other environments).
These matrices are first scaled by the total population to produce a normalized total-contacts matrix, as defined in Eq.~\eqref{pop_sum}.
Next, reciprocity is restored by averaging the total contacts between each pair of age groups (Eq.~\eqref{total}), and the result is divided by the size of the contacting group to yield the full adjusted contact matrix, as described in Eq.~\eqref{sym_contact}.
\begin{figure}[H]
    \centering
    \includegraphics[width=\textwidth]{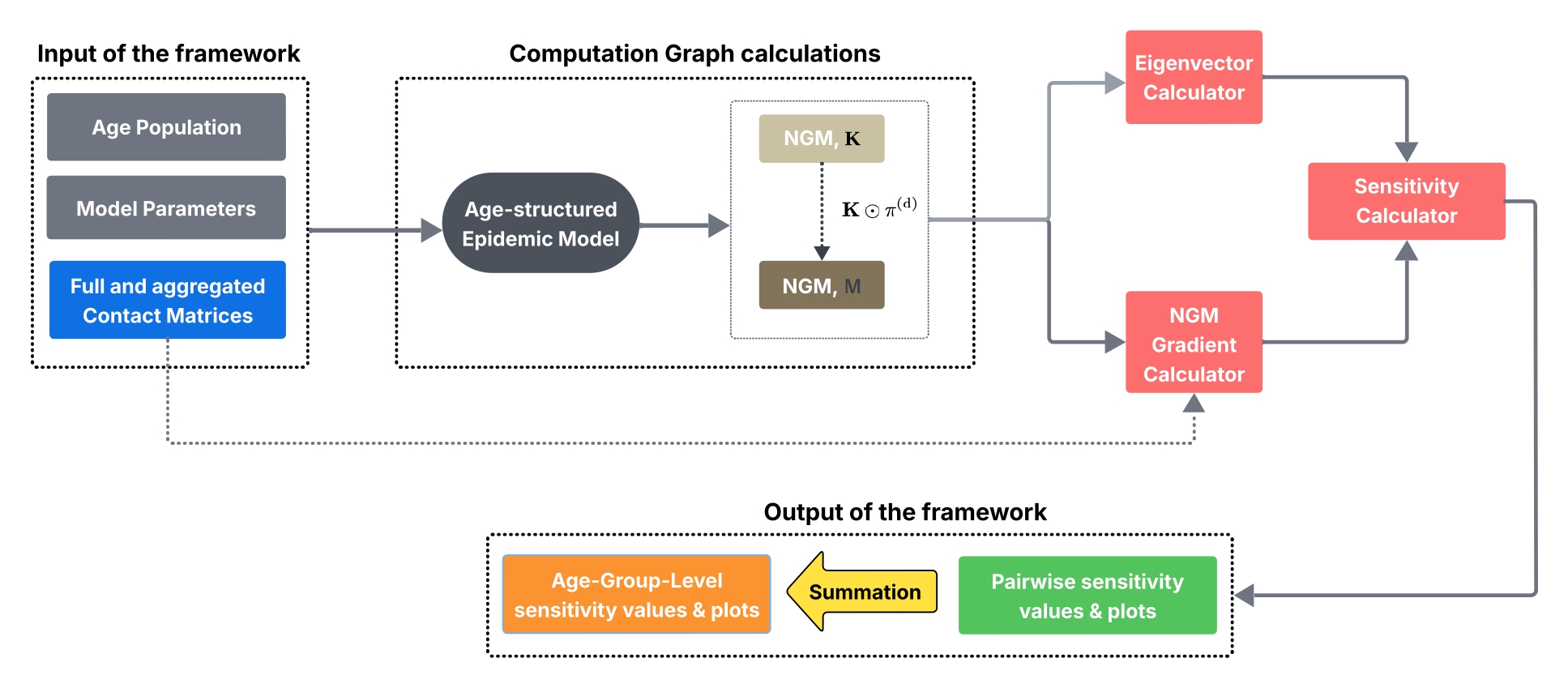}
    \caption{\footnotesize \textbf{Overview of the framework}. This workflow illustrates the pipeline for analyzing the impact of contact patterns on the basic reproduction number, \(\mathcal{R}_0\). It integrates scaled contact matrices from various social settings and employs the Next Generation Matrix methodology to compute \(\mathcal{R}_0\) as the dominant eigenvalue of an epidemic model. The associated eigenvector is then used to derive $n_p$ sensitivity measures, demonstrating how variations in contact inputs affect \(\mathcal{R}_0\). Automatic differentiation, implemented via \texttt{PyTorch}, is used to compute the gradients that underpin the sensitivity measures.}
\label{fig:pipeline}
\end{figure}
The scaled contact matrix \(C^{\mathrm{pop}}\), before these transformations, forms the input to a \textit{computational graph}. This graph propagates the contact data through an age-structured transmission model to compute the next-generation matrix \(\mathbf{K}\), from which the basic reproduction number \(\mathcal{R}_0\) is obtained as its dominant eigenvalue. At this stage, \textit{automatic differentiation}, implemented in \texttt{PyTorch}, is incorporated to compute the gradient of \(\mathcal{R}_0\) with respect to the entries of \(C^{\mathrm{pop}}\). These gradients form the foundation for all subsequent sensitivity calculations.
We use automatic differentiation (AD) not to replace the closed-form eigenvalue sensitivity, but to propagate derivatives through the entire pipeline. For high-dimensional, multi-compartment models (e.g., the Röst model), manual chain-rule derivations are impractical; AD applies the exact chain rule on the computational graph, yielding tractable, reliable gradients, reducing algebraic errors, and making the framework readily extensible across models and outcomes.

The framework flexibly supports both simple and complex age-structured epidemic models for computing \(\mathbf{K}\), making it adaptable to a variety of epidemiological settings. Mortality transition probabilities \(\pi^{(\text{d})}_i\), defined in Eq.~\eqref{eq:death_prob}, can then be applied to transform \(\mathbf{K}\), enabling sensitivity analyses focused specifically on the mortality outcome.

In the \textit{Eigenvector Calculator} (see Fig.~\ref{fig:pipeline}), the framework calculates the eigenvector associated with \(\mathcal{R}_0\) according to Eq.~(\ref{eq: eigen_problem}). The \textit{NGM Gradient Calculator} then uses Eq.~(\ref{eq: r0_cm}) to generate $n_p$ sensitivity values in the \textit{Sensitivity Calculator}, each capturing the impact of variations in specific elements of the contact matrix on \(\mathcal{R}_0\). The sensitivity values are then summed using Eq.~(\ref{eq: cum_sens}) to capture the total influence of each age group on \(\mathcal{R}_0\). The results are presented as heat maps and charts, highlighting the interactions in the contact matrix that most significantly affect disease dynamics. The entire analysis process is automated in Python, using \texttt{PyTorch} for automatic differentiation and \texttt{matplotlib} for visualization.

\subsection{Social Contact Matrices}\label{contacts}
To demonstrate the flexibility of our framework across different age-structured models, we use social contact matrices from two distinct sources representing two geographic contexts: the United Kingdom and Hungary. For the UK, we adopt contact data from Mossong et al.~\cite{mossong2008}, which provides contact matrices for eight European countries based on 7,290 diary-based surveys. These diaries captured the characteristics of 97,904 contacts recorded over a single day, including details such as age, sex, location, duration, frequency, and whether physical contact occurred.  
For Hungary, we use synthetic contact matrices estimated by Prem et al.~\cite{prem}, who used Markov Chain Monte Carlo simulations and additional data sources to produce synthetic contact matrices for various countries based on contact data from Mossong et al.~\cite{mossong2008}. 
These matrices are stratified by setting---home, school, workplace, and other (e.g., public transit, leisure venues)---and reported in five-year age bands.

\begin{figure}[H]
    \centering
    \begin{subfigure}{.18\textwidth}
      \centering
\includegraphics[width=1\linewidth]{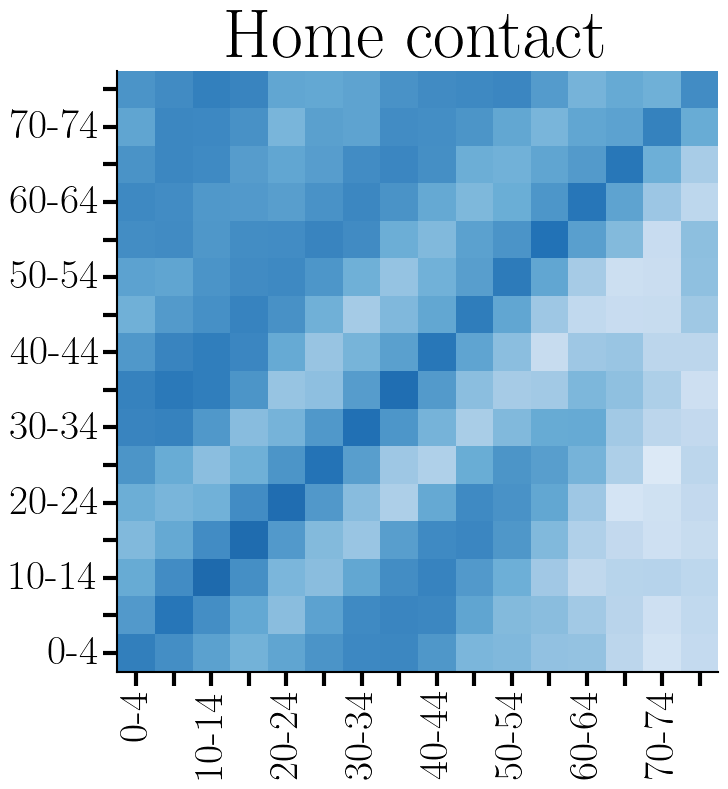}
      \label{fig:home}
    \end{subfigure}%
    \begin{subfigure}{.18\textwidth}
      \centering
\includegraphics[width=1\linewidth]{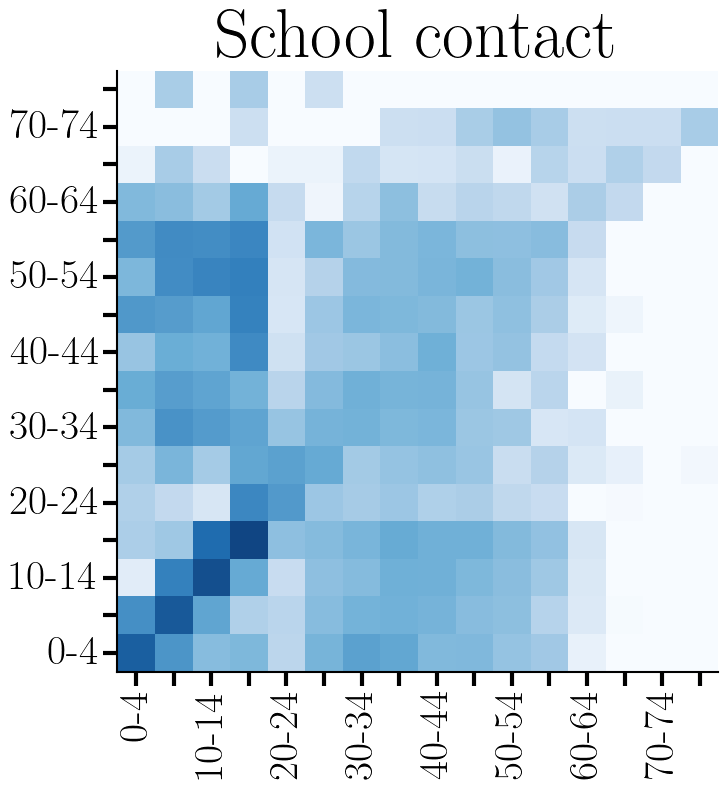}
      \label{fig:school}
    \end{subfigure}
    \begin{subfigure}{.18\textwidth}
      \centering
      \includegraphics[width=1\linewidth]{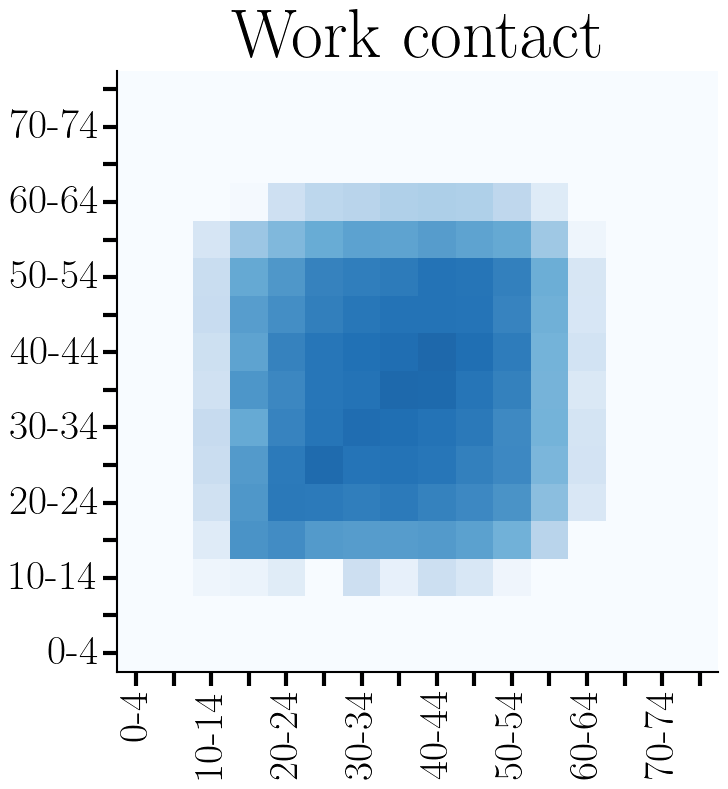}
      \label{fig:work}
    \end{subfigure}
    \begin{subfigure}{.18\textwidth}
      \centering
      \includegraphics[width=1\linewidth]{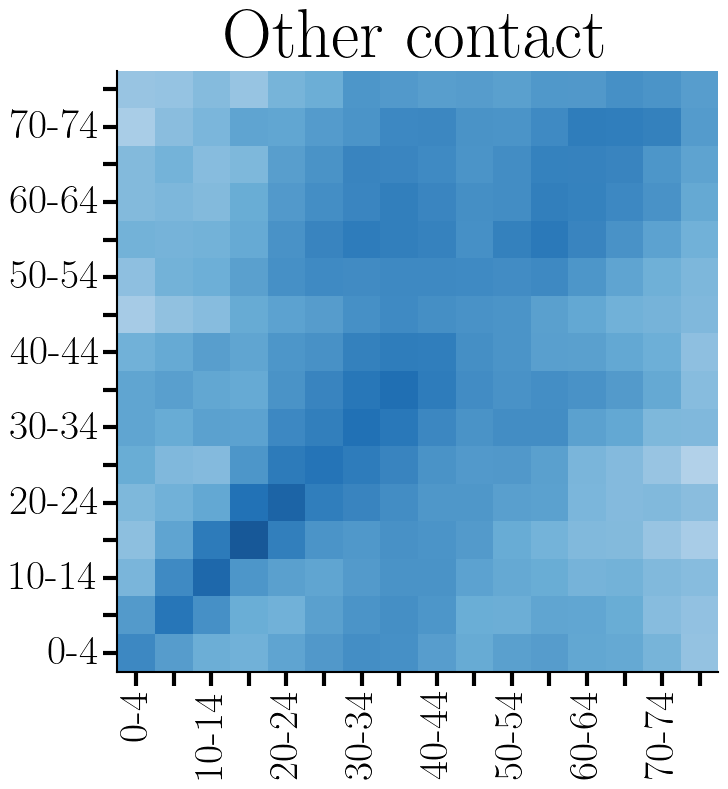}
      \label{fig:other}
    \end{subfigure}
    \begin{subfigure}{.24\textwidth} 
    \centering
    \includegraphics[width=1\linewidth]{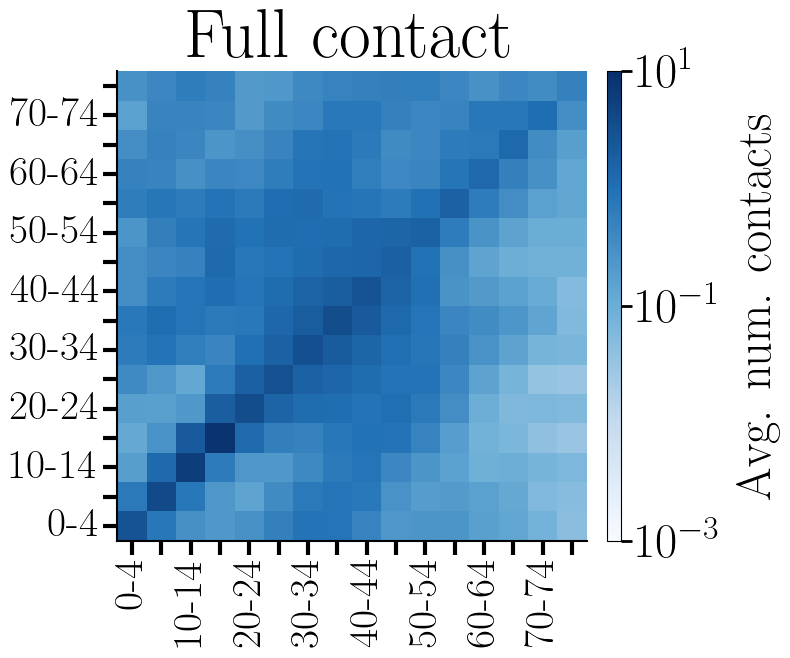}
    \label{fig:cm}
    \end{subfigure} \caption{\footnotesize \textbf{Contact matrices illustrating social interactions in Hungary across different settings: Home, School, Work, and Other locations}. Each heatmap shows the frequency of contacts between age groups (0--4 to 75+ years), where the horizontal axis represents survey participants (respondents), and the vertical axis corresponds to the individuals they reported having contact with (contactees). The color intensity reflects contact frequency, with darker shades indicating higher values. Contact rates are the average number of contacts per person per day. The matrices satisfy population-level reciprocity (Eqs.~\eqref{total}, \eqref{sym_contact}).
    For visual clarity, alternating age group labels are shown along the axes. A consistent logarithmic color scale is applied to all heatmaps to allow direct comparison. The Full contact matrix is computed as a linear combination of the Home, School, Work, and Other matrices, as described in Eq.~(\ref{sym_contact}). Due to the reciprocity condition, only its upper triangular section, containing \( n_p \) unique values, is used for sensitivity analysis. }
    \label{fig:contacts}    
    \end{figure}

To ensure comparability across data sources, we apply a standard procedure to obtain an \textit{adjusted full contact matrix} for each setting and country (details in Appendix~\ref{contact matrix}). The resulting matrices for Hungary are shown in Fig.~\ref{fig:contacts}, and the UK counterparts appear in the left panel of Fig.~\ref{fig:seir}.

\subsection{Sensitivity Measures}\label{concept}
A precise characterization of disease transmission is crucial for developing reliable epidemic models. In age-structured models, transmission dynamics are governed by age-dependent parameters and a social contact matrix. We denote this contact matrix by \( \bar{C} \in \mathbb{R}^{n \times n} \), where \( n \) represents the number of age groups—either \( n = n_a \) for the full-resolution matrix or \( n = n_g \) for its aggregated counterpart. The full-resolution matrix corresponds to the original age groups as provided in the empirical contact data sources (e.g., Prem et al. \cite{prem} or Mossong et al.~\cite{mossong2008}), while the aggregated version combines these groups into broader categories for comparison or to reduce model complexity. The entries of \( \bar{C} \) are positive real numbers, since they represent the frequency of interactions between individuals across age groups. We compute the basic reproduction number \( \mathcal{R}_0 \) using the Next Generation Matrix \( \mathbf{K} \) or the modified $\mathbf{K}$ based on the outcome as detailed in Section~\ref{model}. The value of \( \mathcal{R}_0 \), defined as the dominant eigenvalue of \( \mathbf{K} \), represents the expected number of secondary infections generated by a single infectious individual in a fully susceptible population.

To analyze how changes in contact rates influence \( \mathcal{R}_0 \), we first identify the elements of the contact matrix \( \bar{C} \) that should be varied. Because $\bar{C}$ satisfies population-level reciprocity, each off-diagonal pair $(i,j)$ is determined by a single entry and the population sizes; hence, only the upper (or equivalently, lower) triangular part contains independent values. The total number of these unique contact values is then given by $ n_p = \frac{n(n+1)}{2}.$
We collect these \( n_p \) independent entries into a vector \( \mathbf{c} \in \mathbb{R}^{n_p} \). Each element \( c_p \) of \( \mathbf{c} \) directly corresponds to one of the independent contact rates from the upper triangular part of \( \bar{C} \), and thus represents a positive real value. The gradient of \( \mathcal{R}_0 \) is then computed with respect to the elements in \( \mathbf{c} \).

Using Eq.~(\ref{eq:elasticity}) from Appendix~\ref{sensitivities}, each entry \( S_{ij} \) quantifies the proportional contribution of each element \( k_{ij} \) of \( \mathbf{K} \) to \( \mathcal{R}_0 \) based on the left and right eigenvectors of \( \mathbf{K} \). Following the low-level-parameters approach of \cite{angeli2024acquires}, we treat contact entries as lower-level parameters and therefore compute the sensitivity of \(\mathcal{R}_0\) with respect to \(\mathbf{c}\) by applying the chain rule:
\begin{equation}
    \frac{\partial\mathcal{R}_0}{\partial c_p} = \sum_{i=1}^{n} \sum_{j=1}^{n} S_{ij} \cdot \frac{\partial k_{ij}}{\partial c_p}, \quad
    \label{eq: r0_cm}    
\end{equation}
where $p = 1, \dots, n_p$ and
\begin{itemize}
    \item  \( \frac{\partial k_{ij}}{\partial c_p} \) is a scalar that describes how the entry \(k_{ij}\) of \(\mathbf{K}\) changes with respect to the independent contact matrix elements.    
\end{itemize}
The resulting gradient \( \frac{\partial \mathcal{R}_0}{\partial \mathbf{c}} \in \mathbb{R}^{n_p} \) quantifies the sensitivity of \( \mathcal{R}_0 \) to changes in the $n_p$ independent elements of the contact matrix \( \bar{C} \), representing the effect of perturbations in contact rates on disease transmission.

To summarize the overall impact of contact variations, we reconstruct the sensitivity information from Eq.~\eqref{eq: r0_cm} in matrix form, denoted by 
\(\Sigma = [s_{ij}] \in \mathbb{R}^{n \times n}\). Each entry \(s_{ij}\) equals \(\partial \mathcal{R}_0/\partial c_p\), where \(p\) is the parameter index assigned to the upper-triangular contact pair \((i,j)\); 
the lower triangle is filled by mirroring by setting $s_{ij}=s_{ji}$.
Thus, \(s_{ij}\) quantifies the contribution of contacts between a susceptible individual in group \(i\) and an infectious individual in group \(j\). Diagonal entries capture within-group effects, while off-diagonal entries reflect cross-group contributions.  
Since each column \(j\) of $\mathbf{K}$ represents the distribution of secondary infections produced by an infectious individual in group \(j\), the cumulative transmission impact of group \(j\) is naturally obtained by summing over its column of sensitivities: 
\begin{equation}
    \mathbf{S}_j = \sum_{i=1}^{n} s_{ij}, \quad j = 1, \dots, n.
    \label{eq: cum_sens}
\end{equation}

\subsection{Age-structured Epidemic Models}\label{model}
To illustrate our sensitivity analysis framework, we apply it to two age-structured epidemic models of varying complexity. Rather than constructing new models, we use two published compartmental models that incorporate age-specific contact matrices and transmission mechanisms: one by Pitman et al. (2012) \cite{pitman2012estimating}, and the other by Röst et al.~\cite{Rost}. The objective of these case studies is to demonstrate that our eigenvector-based sensitivity framework can be applied consistently across models of different complexity, ranging from a relatively simple SEIRS-type system to a detailed multi-compartment COVID-19 model, and to highlight how demographic structures, contact matrices, and model assumptions influence the resulting sensitivity patterns. The Pitman et al.\ model serves as a classical influenza example for illustrating the methodology, whereas the Röst et al.\ model provides the opportunity to explore mortality-related outcomes in the context of COVID-19.

In both models, we compute the basic reproduction number \( \mathcal{R}_0 \) using the Next Generation Matrix (NGM) framework introduced by Diekmann et al.~\cite{diekmann}, where \( \mathcal{R}_0 \) is defined as the spectral radius of the NGM, \( \mathbf{K} \). This approach involves linearizing the infection-related compartments of the system around the disease-free equilibrium. The resulting linear system is decomposed into two matrices: the transmission matrix \(F\), which describes the rate at which new infections are produced, and the transition matrix \(V\), which governs the movement between infectious compartments. We construct these matrices for each model in Sections~\ref{pitman_ngm} and~\ref{rost_ngm}.

\subsubsection{Model 1: Pitman et al.}\label{pitman_ngm}

The first model is an SEIRS (Susceptible–Exposed–Infectious–Recovered–Susceptible) framework developed to simulate the transmission of influenza, incorporating age-specific contact patterns in England and Wales \cite{pitman2012estimating}. The model assumes uniform susceptibility across all age groups and sets the baseline reproduction number \( \overline{\mathcal{R}}_0 \) to 1.8. The system of governing equations is provided in Appendix~\ref{sec:pitman}. 

The infected state vector \( X(t) \) comprises \( n_s \) compartments, \(E_i\) and \(I_i\), for each age group \( i = 1, \dots, n \). New infections enter only the \(E_i\) compartments, while individuals in \(I_i\) contribute to transmission but do not directly receive new inflow.
The transmission matrix \( F \in \mathbb{R}^{(n \cdot n_s) \times (n \cdot n_s)} \) is sparse, as new infections occur only in the \( E_i \) compartments through contact with infectious individuals \( I_j \). It is constructed as a block matrix with sub-blocks \(F_{j,i}\), each representing transmission from infectious individuals in age group \(j\) to susceptibles in age group \(i\):
\[
F_{j,i} =
\beta \cdot \overline{c}_{j,i} \cdot
\begin{bmatrix}
0 & 1 \\
0 & 0
\end{bmatrix}.
\]

The progression matrix \( V \) is block-diagonal, with each diagonal block \( V_{i,i} \in \mathbb{R}^{n_s \times n_s} \) defined by:
\[
V_{i,i} =
\begin{bmatrix}
-\alpha & 0 \\
\alpha & -\gamma
\end{bmatrix},
\]
where \( \alpha \) is the rate of progression from exposed to infectious, and \( \gamma \) is the recovery rate. The negative diagonal entries capture the outflow from the exposed compartment (\(E_i \to I_i\)) and from the infectious compartment (\(I_i \to R_i\)).

To reduce the full $(n \cdot n_s)\times(n \cdot n_s)$ system to an $(n\times n)$ NGM, we define the selection matrix \( \mathcal{E} \in \mathbb{R}^{n \times (n \cdot n_s)} \). 
The \(i\)-th row of \( \mathcal{E} \) selects the newly infected state \(E_i\) from the local pair \([E_i,\, I_i]\) (i.e., it has the block \([1\ 0]\) at the columns of age group \(i\)) and zeros elsewhere; \(I_i\) is not selected.
The NGM is then computed as:
\begin{equation}
\mathbf{K} = -\,\mathcal{E} \cdot F V^{-1} \cdot \mathcal{E}^\top, 
\quad \text{with } \mathbf{K} \in \mathbb{R}^{n \times n}, 
\quad \mathcal{R}_0 = \rho(\mathbf{K}).
    \label{eq: ngm}
\end{equation}

\subsubsection{Model 2: Röst et al.}\label{rost_ngm}

The Röst et al.\ model (section \ref{sec:rost}) includes \(n = 16\) age groups, each with \(n_s = 9\) infectious compartments: \(L^{(1)}, L^{(2)}, I^{(p)}, I^{(a,1)}, I^{(a,2)}, I^{(a,3)}, I^{(s,1)}, I^{(s,2)}, I^{(s,3)}\). The infected state vector \(X(t)\) therefore contains all of these compartments for each age group.
Consequently, the full transmission matrix has dimension \(F \in \mathbb{R}^{(n \cdot n_s) \times (n \cdot n_s)}\).
Each entry in this matrix reflects contributions governed by the contact rate \(\overline{c}_{j,i}\), the age-specific susceptibility \(\sigma_i\), and the transmission rate \(\beta\), as described in the original formulation~\cite{Rost}. The matrix is composed of \(n \times n\) blocks \(F_{j,i}\), where each block has shape \(n_s \times n_s\). Importantly, only the first row of each block is non-zero, since all new infections enter the first latent compartment \(L^{(1)}_i\). The nonzero entries capture contributions from pre-symptomatic (\(I^{(p)}_j\)), asymptomatic (\(I^{(a,m)}_j\)), and symptomatic (\(I^{(s,m)}_j\)) individuals with reduced transmissibility for asymptomatic cases (\(\inf^{(a)} = 0.5\)). 

The progression matrix \(V \in \mathbb{R}^{(n \cdot n_s) \times (n \cdot n_s)}\) is block-diagonal, with each diagonal block \( V_{i,i} \in \mathbb{R}^{n_s \times n_s} \) encoding transitions within the infectious compartments of age group \(i\), including progression from latent to pre-symptomatic, branching into asymptomatic or symptomatic paths, and recovery. To reduce the full system to an \((n \times n)\) NGM, we define a selection matrix \(\mathcal{E} \in \mathbb{R}^{n \times (n \cdot n_s)}\). The \(i\)-th row of \(\mathcal{E}\) selects the newly infected compartment \(L^{(1)}_i\) from the set of infectious states for age group \(i\). The reduced NGM is then computed as in Eq.~(\ref{eq: ngm}).

As seen in Eq.~(\ref{eq: rost}), this model is sufficiently complex to include various compartments for hospitalization and death. While it supports analysis of health outcomes such as \textit{hospitalization} (\(I^{(h)}\)), \textit{ICU} admission (\(I^{(c)}\)), and \textit{death} (\(D\)), our focus here is solely on mortality. Accordingly, we consider only the probability of death when modifying \(\mathbf{K}\).

The age-specific probability of death is derived from the model’s compartmental parameters and reflects the conditional pathway from presymptomatic infection to a fatal outcome:
\begin{equation}
\pi_i^{(\text{d})}
= \Pr\!\left(I^{(p)}_i \rightarrow D_i\right)
= (1 - p_i)\, h_i\, \xi_i\, \mu_i,
\label{eq:death_prob}
\end{equation}
where $\pi_i^{(\text{d})}$ is the probability that an individual in age group $i$ progresses from the presymptomatic compartment $I^{(p)}_i$ to death.

To explore \emph{mortality-related} sensitivities in the epidemic process, we define a mortality-weighted matrix $\mathbf{M} = [m_{ij}]$ by scaling the entries of $\mathbf{K}$ with the age-specific probabilities of death:
\begin{equation}
m_{ij} \;=\; k_{ij}\,\pi_j^{(\text{d})}.
\label{eq:mortality_weighted_matrix}
\end{equation}
Unlike the transmission analysis in which $\mathcal{R}_0$ is a well-defined epidemiological target, the mortality-weighted analysis does not introduce a new explicit target quantity; rather, it yields a conceptual target that captures the effectiveness of lethality (i.e., transmission weighted by fatality risk). Importantly, this measure should not be interpreted as the sensitivity of the cumulative mortality trajectory, but as a heuristic indicator of which contact pathways most strongly influence mortality-weighted transmission.
This construction comes with important limitations. By scaling the NGM in this way, deaths are effectively assumed to occur instantaneously after infection, overlooking temporal delays between infection, hospitalization, and fatal outcomes. Therefore, $\mathbf{M}$ should not be regarded as a new NGM, and its spectral radius is not a mortality-specific reproduction number. Instead, the role of $\mathbf{M}$ is heuristic: it reweights transmission pathways according to their mortality consequences, and the associated left and right eigenvectors highlight which age-specific contacts contribute most to fatal outcomes. In this way, the sensitivity framework of Eq.~\eqref{eq: r0_cm} is adapted to emphasize mortality-related effects rather than transmission itself.

The choice between infection-based and mortality-weighted sensitivities ultimately depends on the epidemiological context and intervention goals: reducing transmission is critical when epidemic control and eradication are priorities, while focusing on mortality-weighted measures is more relevant in settings where severe outcomes dominate the public health burden. This trade-off is analogous to vaccination strategies, where prioritizing younger high-contact groups curbs spread, while targeting older groups reduces deaths.

\section{Results}\label{results}
In this section, we present the results of our framework for quantifying the sensitivity of transmission dynamics to age-specific contact patterns. The contact matrices serve as key inputs to the models and are the primary focus of the sensitivity analysis. As a case study, we apply the method to the \emph{influenza} SEIR model of Pitman et al. (Section~\ref{sec:pitman}) using age-structured contact data from the UK.
To validate the framework in a different epidemiological context, we repeat the analysis using 
Röst et al. model described in Section~\ref{sec:rost}, using the Hungarian contact data shown in Fig.~\ref{fig:contacts}. 
In addition to the full-resolution analysis, we also investigate the effect of contact matrix aggregation on the sensitivity results using the Röst model. 
Model parameter values and assumptions are taken directly from the original studies \cite{pitman2012estimating, Rost}.

\subsection{Age-structured \emph{Influenza} SEIR model by Pitman et al}
\begin{figure}[H]
\small
\centering
\includegraphics[width=0.31\textwidth]{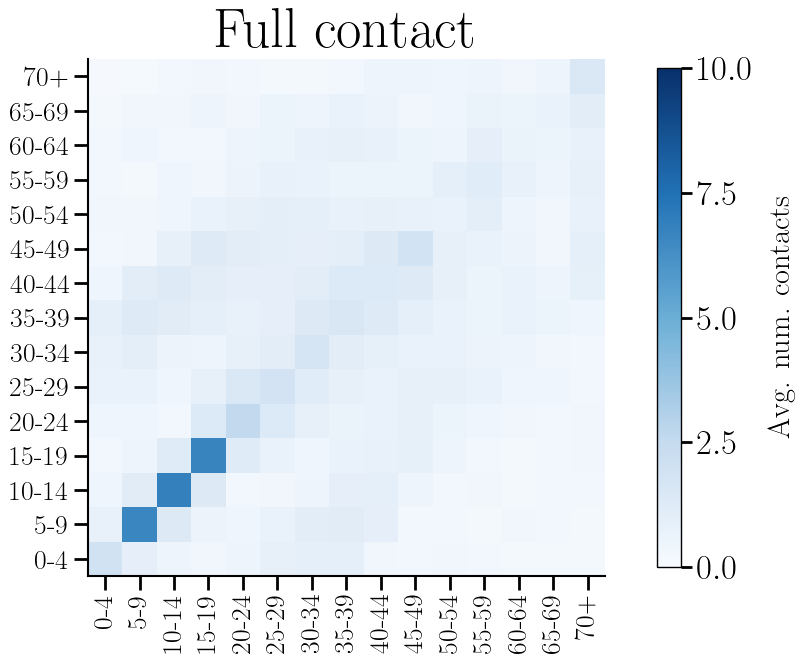}
\includegraphics[width=0.33\textwidth]{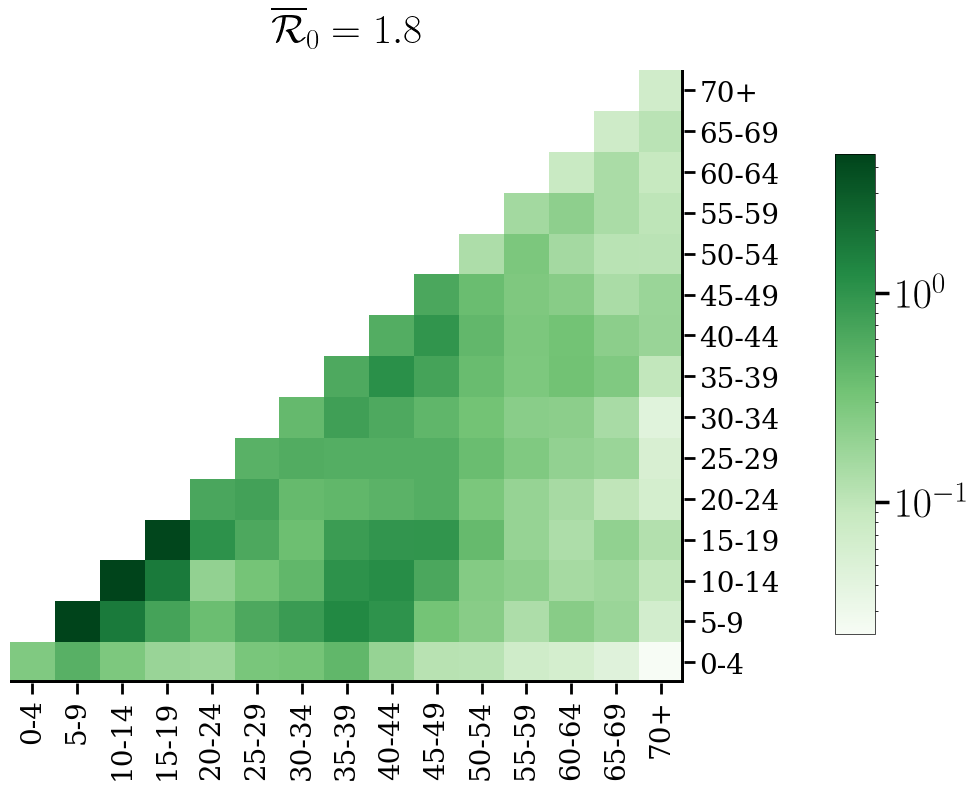}
\includegraphics[width=0.33\textwidth, height=0.18\textheight]{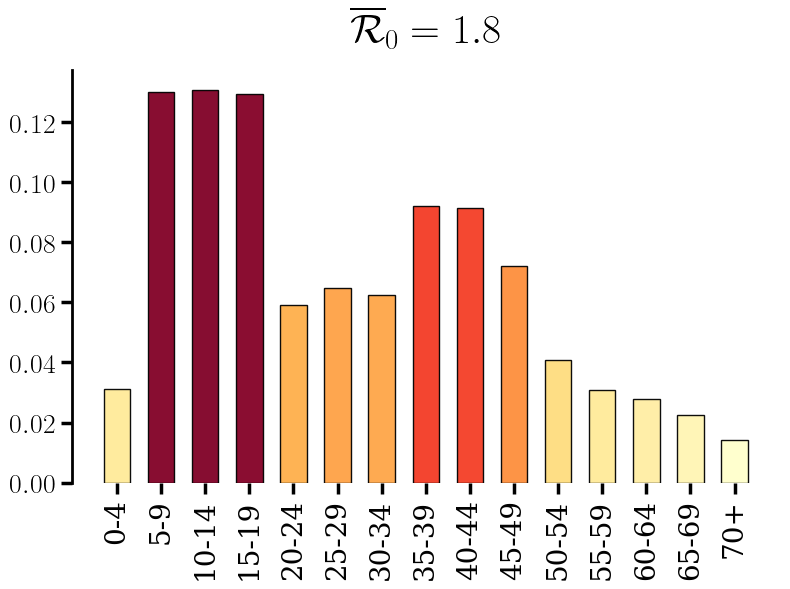}
\caption{\small The left panel shows the age-specific social contact matrix from Mossong et al.~\cite{mossong2008}, reflecting contact frequencies between age groups. 
The middle panel displays pairwise sensitivity scores, indicating how changes in age-group interactions affect \(\mathcal{R}_0\). 
The right panel shows the group-level sensitivities, summarizing the total contribution of each age group to transmission potential. All results shown pertain to an \emph{influenza} SEIR model (Pitman et al.).}

\label{fig:seir}
\end{figure}

This SEIR model provides a straightforward example to demonstrate the framework, using age-specific mixing patterns from England and Wales (Fig.~\ref{fig:seir}, left panel). The model quantifies the sensitivity of the basic reproduction number \(\mathcal{R}_0\) to contact rates, although it does not allow exploration beyond transmission-related outputs.

The middle panel of Fig.~\ref{fig:seir} shows the sensitivity of \(\mathcal{R}_0\) to pairwise age group interactions. Notably, contacts among individuals aged 5--19 have the highest influence, suggesting that transmission dynamics are particularly sensitive to interactions in this age range. 
The right panel presents age-group-level  sensitivity scores summed across contact pairs. It reveals that both younger individuals (ages 5--19) and middle-aged adults (ages 35--49) make the largest contributions to variations in \(\mathcal{R}_0\). 

\subsection{Sensitivity Analysis Using the Röst et al.~COVID-19 Model}

To assess the applicability of our framework to more complex epidemiological settings, we apply it to the age-structured COVID-19 model developed by Röst et al.~\cite{Rost}, which was calibrated for the Hungarian population. This model features 15 compartments, incorporating Erlang-distributed latent and infectious periods, distinct hospitalization stages, and fatality tracking. It uses synthetic age-specific contact matrices for Hungary, derived from Prem et al.~\cite{prem} and illustrated in Fig.~\ref{fig:contacts}.
Model parameters, including age-dependent recovery, transition rates, and susceptibility, are taken from the original study. The basic reproduction number is fixed at \(\mathcal{R}_0 = 1.8\).

\begin{figure}[H]
\centering
\includegraphics[width=0.49\textwidth]{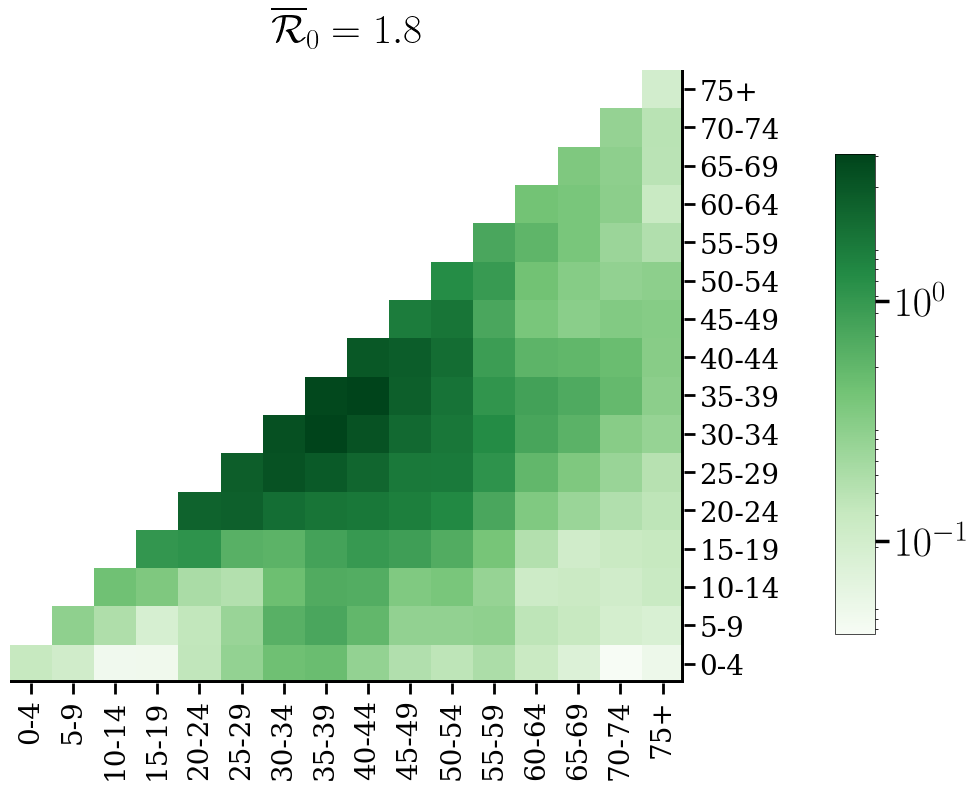}
\includegraphics[width=0.49\textwidth]{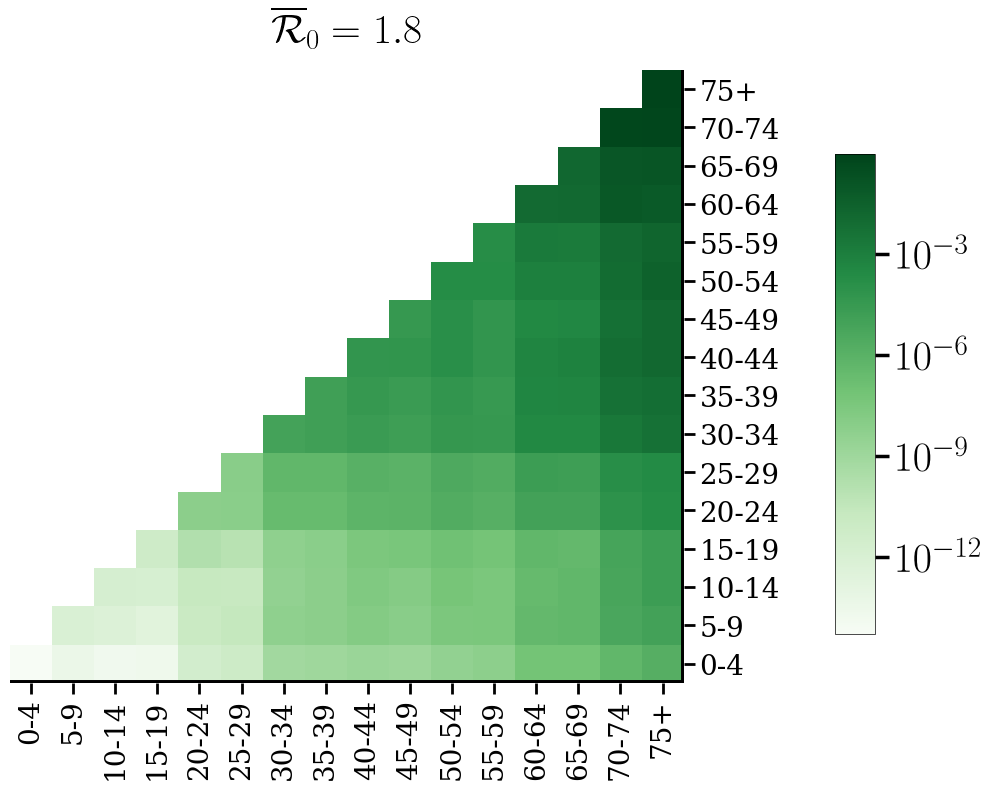}
\caption{\small 
Sensitivity values are computed assuming a 50\% reduction in susceptibility for the 0--19 age group.  (Left) Sensitivity of the basic reproduction number \(\mathcal{R}_0\) to contact perturbations. 
(Right) Sensitivity of \textit{mortality} outcomes. Darker green shades indicate higher sensitivity values, meaning greater influence on the corresponding model output; lighter shades indicate lower sensitivity.}
\label{fig: sens_rost}
\end{figure}

We compute the sensitivity of two key epidemiological outcomes, \(\mathcal{R}_0\) and \textit{cumulative mortality}, with respect to perturbations in contact rates. 
Here, cumulative mortality refers to the total number of deaths predicted by the model over the full course of the epidemic, aggregated across all age groups.
Results are summarized in Fig.~\ref{fig: sens_rost}. The left panel shows that contacts among individuals aged 20–54 exert the strongest influence on \(\mathcal{R}_0\), highlighting their central role in transmission dynamics. However, \textit{mortality} (right panel) sensitivities shift toward contacts involving older adults, particularly those aged 60 and above.

\begin{figure}[H]
\centering
\includegraphics[width=0.49\textwidth]{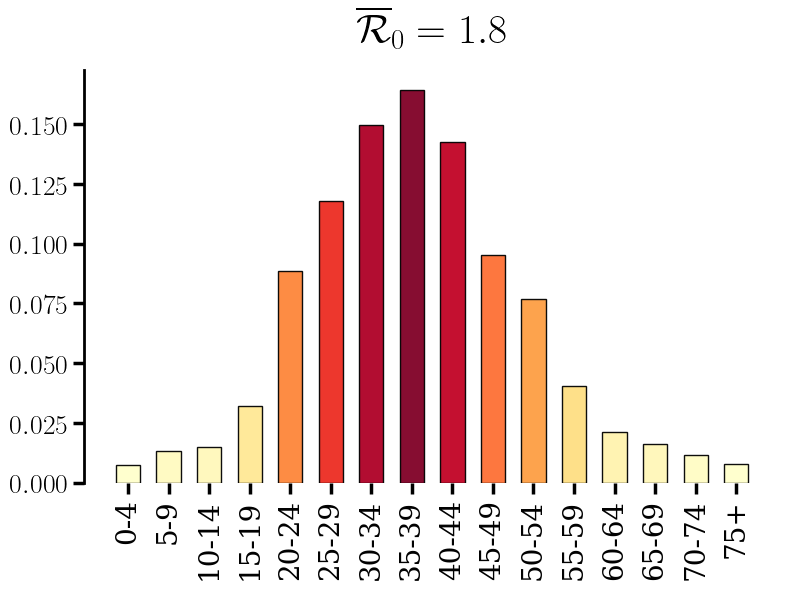}
\includegraphics[width=0.49\textwidth]{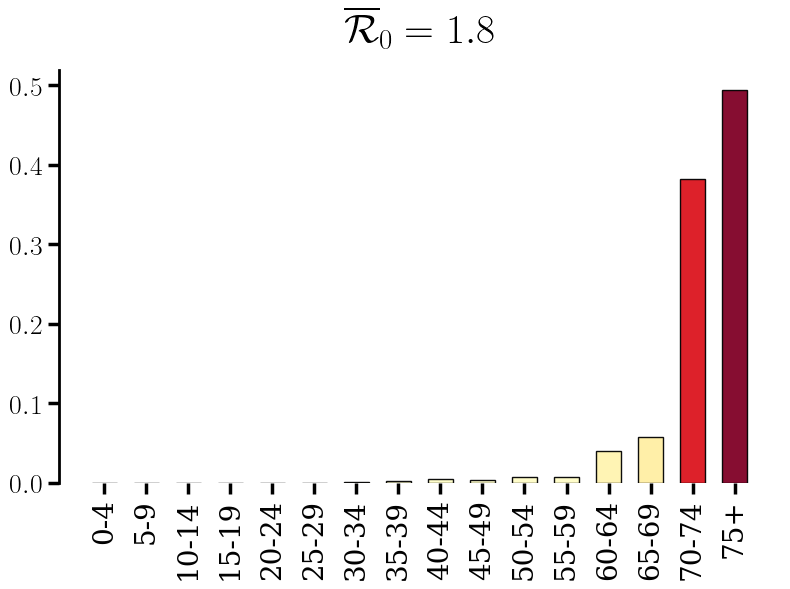}
\caption{\small
\textbf{Group-Level Sensitivity Analysis.}  
(Left) Group-level sensitivity values for \(\mathcal{R}_0\), reflecting the total influence of each age group on transmission dynamics.  
(Right) Group-level sensitivities for cumulative mortality.  
Bar heights represent overall impact, with darker red shades indicating higher sensitivity.}
\label{fig: cum_rost}
\end{figure}

To better understand which age groups most influence model outcomes, we compute \textit{age-group-level sensitivities} by summing the pairwise sensitivity values associated with each age group (as defined in Eq.~\eqref{eq: cum_sens}). These results are shown in Fig.~\ref{fig: cum_rost}.
Age-group sensitivity for \(\mathcal{R}_0\) peaks in the 20–54 age range, indicating that improved estimation of contact patterns and transmission parameters in this demographic could substantially enhance model fidelity. Conversely, the sensitivities for \textit{mortality} are concentrated among individuals aged 70 and above, underscoring the critical importance of accurate clinical parameterization and data collection for these older age groups. These insights can help prioritize efforts to refine age-specific contact matrices and outcome probabilities in the most impactful segments of the population, thereby improving the overall reliability and policy relevance of age-structured epidemic models.

\subsubsection{Sensitivity Analysis Using the Model with Aggregated Age Groups}
\begin{figure}[H]
\small
\centering
\includegraphics[width=0.49\textwidth]{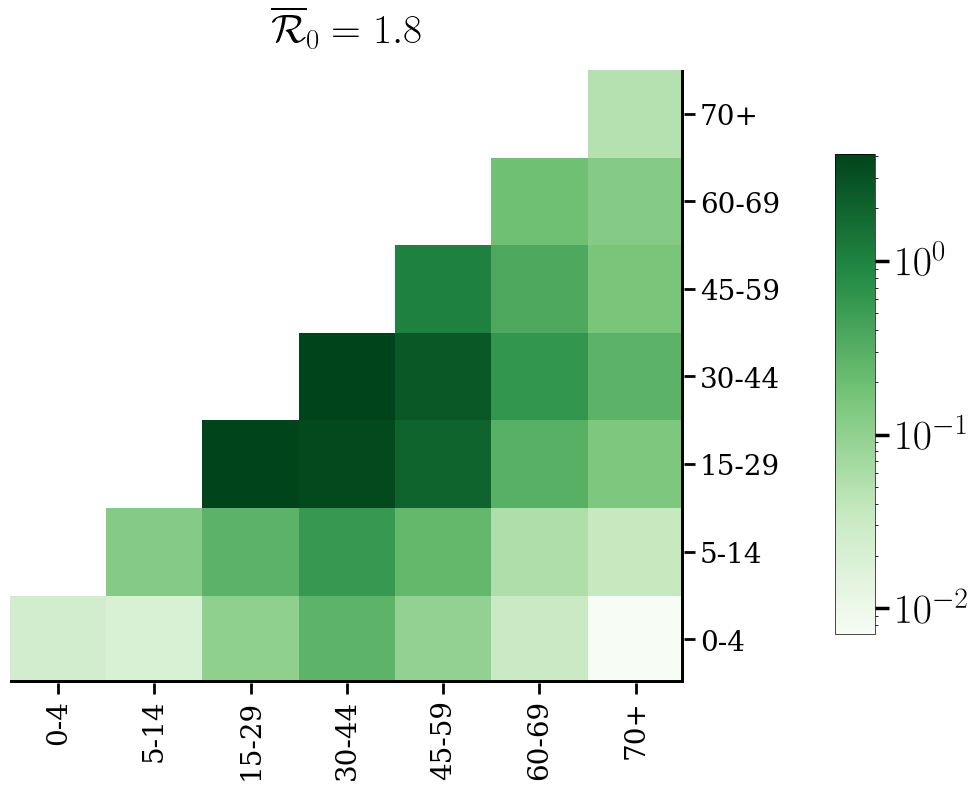}
\includegraphics[width=0.49\textwidth]{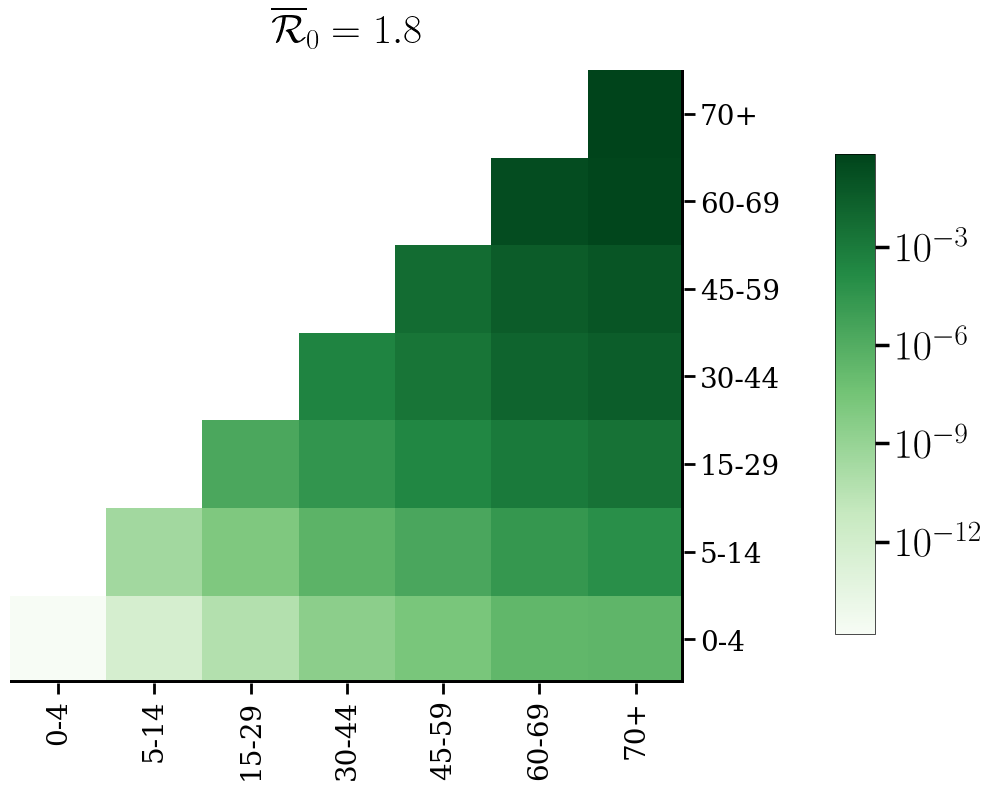}
\includegraphics[width=0.49\textwidth]{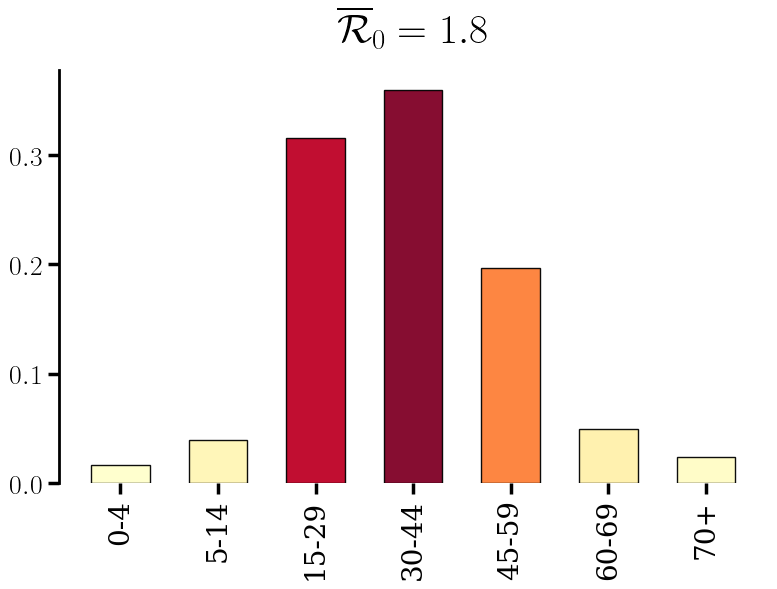}
\includegraphics[width=0.49\textwidth]{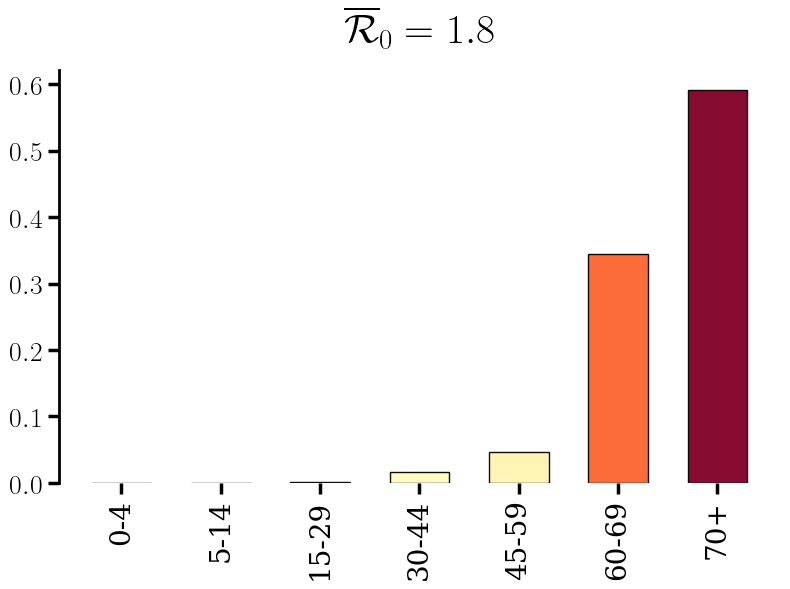}
\caption{\small
\textbf{Sensitivity Analysis with Aggregated Contact Matrix.} 
Top row: Pairwise sensitivity values for \(\mathcal{R}_0\) (left), and \textit{mortality} outcomes (right), assuming a 50\% reduction in susceptibility for the 0–14 age group. 
Bottom row: Corresponding group-level sensitivities for each age group. 
As in the full-resolution case, transmission sensitivities are dominated by contacts involving ages 15–44, while \textit{mortality} sensitivities remain highest for older adults (60+).}
\label{fig: rost_agg}
\end{figure}

To assess the influence of age aggregation on sensitivity results, we applied the framework to a contact matrix aggregated from 16 to 7 age groups, as described in Eq.~\eqref{eq: agg_matrices}. The resulting sensitivities for transmission and clinical outcomes are presented in Fig.~\ref{fig: rost_agg}. Compared to the full-resolution analysis, the sensitivity for \(\mathcal{R}_0\) now shifts slightly toward a broader age group, 15–59 years, indicating that aggregation blends the influence of younger and middle-aged groups.
For the clinical outcomes of \textit{mortality}, the highest sensitivities remain concentrated in the oldest age categories. However, with aggregation, the peak shifts from 70+ in the full model (Fig.~\ref{fig: cum_rost}) to 60+ in the coarser model. 

\section{Discussion}

This study introduces a generalizable sensitivity analysis framework that integrates age-structured social contact data into compartmental epidemic models to assess the impact of contact heterogeneity on disease outcomes. The core objective is to quantify how variations in contact patterns between age groups influence key epidemiological quantities, such as the basic reproduction number \(\mathcal{R}_0\), hospitalizations, ICU admissions, and mortality.

A distinguishing feature of our approach is its ability to decompose sensitivities across the full contact matrix and sum them to identify age groups with the most influential contact patterns. Building on ideas from \cite{angeli2024acquires}, we incorporated cumulative sensitivity metrics to highlight which age groups contribute most substantially to model outputs, guiding targeted data collection and intervention design. To address the known issue that arbitrary age bin definitions can distort sensitivity analyses, we further extended the framework to support analyses over both fine- and coarse-grained age structures. 

A significant implication of this framework is its ability to inform data collection. Because contact matrices are difficult to estimate accurately, especially in underrepresented age groups, our method helps prioritize empirical efforts by pinpointing which contact interactions carry the greatest influence or uncertainty. This focus enables more efficient and impactful refinement of contact data used in infectious disease modeling.
At the same time, two important interpretive limitations should be noted. First, contact rates between age groups are not truly independent quantities. In practice, a change in contacts between groups \(i\) and \(j\) will typically influence other entries of the contact matrix, reflecting constraints from household composition, social networks, and cultural factors. Our framework, like most perturbation analyses, treats contacts as independent for tractability. The resulting sensitivities should therefore be understood as hypothetical responses to isolated perturbations rather than literal predictions of feasible behavioral change. To mitigate this issue, we emphasize cumulative sensitivities at the age-group level, which provide more robust insights than pair-wise sensitivities.

Second, the sensitivity patterns depend on the choice of age partition. Our results with aggregated contact matrices demonstrate that grouping can shift the apparent drivers of transmission and mortality (e.g., transmission sensitivities shifting from 20–54 to 15–59, mortality from 70+ to 60+). This dependence is inherent to age-structured modeling and highlights the importance of carefully selecting age bins. In practice, partitions should balance epidemiological relevance (e.g., school-age, working-age, elderly), policy needs, and data availability. Sensitivity results should therefore be interpreted in light of the chosen age structure, and complementary analyses using alternative partitions can help assess robustness.

The framework offers several key advantages. First, its deterministic nature provides reproducibility and computational efficiency, making it well-suited for rapid policy analysis. Second, the method is outcome-flexible, enabling sensitivity assessments not only for transmission metrics like \(\mathcal{R}_0\), but also for downstream epidemiological outcomes such as hospitalization, ICU occupancy, or mortality, within the same model infrastructure. This versatility supports localized, age-specific public health strategies across diverse model types and settings.
Despite these strengths, the framework also has other limitations. It is based on point estimates and does not currently incorporate statistical uncertainty, which constrains its interpretability in probabilistic terms. The underlying sensitivity analysis assumes local linearity around a baseline scenario and may not fully capture dynamics in highly nonlinear or unstable regimes. Another limitation is the assumption of static contact matrices, which restricts the framework's ability to account for temporal variation in contact behavior due to evolving policies, interventions, or social norms. Nonetheless, the framework is readily extendable to accommodate time-varying contact matrices. Provided that contact data are available across distinct time intervals, the sensitivity analysis can be applied independently to each temporal slice over specified periods as in \cite{angeli2024acquires}. Future work will also aim to embed uncertainty quantification techniques, such as bootstrapping or Bayesian inference, to produce confidence intervals for sensitivity estimates.

\backmatter


\bmhead{Acknowledgements}
On behalf of the RRF-2.3.1-21-2022-00006 project, we are grateful for the possibility of using HUN-REN Cloud (see \cite{heder2022past}, \url{https://science-cloud.hu/})  which helped us achieve the results published in this paper.

\section*{Declarations}

The authors declare that they have no known competing financial interests or personal relationships that could have appeared to influence the work reported in this paper.

\begin{itemize}
\item Funding

This project was supported by the National Laboratory for Health Security Program (RRF-2.3.1-21-2022-00006) and the NKFIH KKP 129877, both funded by the Ministry of Innovation and Technology of Hungary through the National Research, Development, and Innovation Fund. The views and opinions expressed in this work are solely those of the author(s) and do not necessarily reflect those of the National Laboratory for Health Security Program or the NKFIH KKP 129877 project. Neither entity assumes responsibility for these views.

\item Conflict of interest/Competing interests 

The authors declare that they have no competing interests.

\item Ethics approval and consent to participate

Not applicable.

\item Consent for publication

Not applicable.

\item Data and Code Availability

All the code and data required to reproduce our results are available in the repository \cite{code}.

\item Author contribution

\textbf{Evans Kiptoo Korir:} Contributed to methodology, conceptualization, drafting the original manuscript, reviewing and editing, software development, and visualization.

\textbf{Zsolt Vizi}: Responsible for data curation, conceptual framework, formal analysis, methodology, project management, software development, visualization, manuscript review and editing, and supervision. 

\end{itemize}

\noindent

\begin{appendices}

\section{Social Contact Matrices}\label{contact matrix}
We use Hungarian contact data estimated in \cite{prem} to describe the contact patterns applied in this study.
The contact matrices for different settings---home ($\widetilde{C}^\text{H}$), school ($\widetilde{C}^\text{S}$), work ($\widetilde{C}^\text{W}$), and other environments ($\widetilde{C}^\text{O}$)---are combined to form the full contact matrix:
\begin{equation*}
\widetilde{C} = \widetilde{C}^\text{H} + \widetilde{C}^\text{S} + \widetilde{C}^\text{W} + \widetilde{C}^\text{O}.
\end{equation*}
This full contact matrix $\widetilde{C} \in \mathbb{R}^{n_a \times n_a}$ consists of elements $\widetilde{c}_{i,j}$, representing the average number of daily contacts reported by individuals in age group $i$ with those in age group $j$. 
Due to sampling variability, the matrix is often asymmetric and does not satisfy reciprocity.
In theory, contact matrices should satisfy the \emph{reciprocity condition}
\[
    \widetilde{c}_{i,j} N_i = \widetilde{c}_{j,i} N_j,
\]
where $N_i$ and $N_j$ are the population sizes of age groups $i$ and $j$. 
To enforce reciprocity, we apply the adjustment method of Knipl et al.~\cite{knipl}. 
First, we construct the total contact matrix
\begin{equation}
    c^{\text{total}}_{i,j} = \widetilde{c}_{i,j} N_i.
\end{equation}
This matrix is generally not symmetric. We then enforce symmetry at the level of total contacts by averaging:
\begin{equation}
    \tilde{c}^{\text{total}}_{i,j} = \frac{\widetilde{c}_{i,j} N_i + \widetilde{c}_{j,i} N_j}{2}.
\label{total}
\end{equation}
By construction, $\tilde{C}^{\text{total}} = [\tilde{c}^{\text{total}}_{i,j}]$ is symmetric, and when rescaled to per-capita form (Eq.~\eqref{sym_contact}), the resulting matrix satisfies reciprocity.
To obtain the per-capita contact matrix used in transmission modeling, we divide by the size of the contacting age group:
\begin{equation}
    \overline{c}_{i,j} = \frac{\tilde{c}^{\text{total}}_{i,j}}{N_i}.
\label{sym_contact}
\end{equation}
The resulting $\overline{C} = [\overline{c}_{i,j}]$ specifies, for each individual in age group $i$, the average number of contacts with individuals in age group $j$. 
For sensitivity analysis, we also introduce a population-scaled version
\begin{equation}
    c^{\mathrm{pop}}_{i,j} = \frac{\tilde{c}^{\text{total}}_{i,j}}{\sum_{k=1}^{n_a} N_k},
\label{pop_sum}
\end{equation}
which is not required in the classical definition of contact matrices, but is used in our computation graph. This normalized form ensures that contact entries are scaled consistently with the total population, serving as the foundation for gradient-based sensitivity calculations.

The adjusted contact matrices $\overline{C} \in \mathbb{R}^{n_a \times n_a}$, computed via Eq.~\eqref{sym_contact}, are visualized in Fig.~\ref{fig:contacts}. Because of the reciprocity condition, the computational graph processes only the upper (or equivalently, lower) triangular entries of $C^{\mathrm{pop}}$, which amounts to $n_p$ unique values. These values sufficiently describe all pairwise contact interactions and serve as inputs for sensitivity analysis.

To examine how the level of age resolution influences model sensitivity, we also construct aggregated versions of the contact matrix from the full $\widetilde{C} \in \mathbb{R}^{n_a \times n_a}$, following the aggregation approach described in \cite{korir}.
Let $G_i$ denote the set of original age indices grouped into the $i$-th new age bin, with $i = 1, \dots, n_g$ and $n_g < n_a$. 
The population in the new age bin is defined as $\mathbf{N}_i = \sum_{m \in G_i} N_m$. 
The aggregated contact matrix $\mathbf{\widetilde{C}} \in \mathbb{R}^{n_g \times n_g}$ is then computed by taking a population-weighted average of contact rates between all combinations of original age groups:
\begin{equation}
    \mathbf{\widetilde{c}}_{i,j} = \frac{1}{\mathbf{N}_i} \sum_{(m, m') \in G_i \times G_j} \widetilde{c}_{m,m'} N_m.
\label{eq: agg_matrices}
\end{equation}
This aggregated matrix is subsequently processed using the same reciprocity-restoring and normalization procedures as the full contact matrix, yielding the final per-capita aggregated matrix $\mathbf{\overline{C}} \in \mathbb{R}^{n_g \times n_g}$ used in sensitivity analysis.

\section{General Sensitivity Measures} \label{sensitivities}

For a given matrix \(\mathbf{K}\), the eigenvalue \(\lambda\) is associated with the left and right eigenvectors, which provide crucial insight into the relationship between \(\mathbf{K}\) and \(\lambda\). These eigenvectors satisfy the following equations:

\begin{equation}
\mathbf{K}\mathbf{w} = \lambda \mathbf{w}
\label{eq: eigen_problem}
\end{equation}
\begin{equation*}
\mathbf{v}^\mathrm{T}\mathbf{K} = \lambda \mathbf{v}^\mathrm{T}.
\end{equation*}
Since $\mathbf{K}$ is a strictly positive matrix, its dominant eigenvalue $\mathcal{R}_0$ is real, positive, and algebraically simple. It admits a corresponding right eigenvector $\mathbf{w}$ and left eigenvector $\mathbf{v}$, both with strictly positive real components, satisfying:
\[
\mathbf{K} \mathbf{w} = \mathcal{R}_0 \mathbf{w}, \quad \mathbf{v}^\mathrm{T} \mathbf{K} = \mathcal{R}_0 \mathbf{v}^\mathrm{T}.
\]
Thus, the leading eigenvalue $\lambda = \mathcal{R}_0$ governs the asymptotic growth rate of the system.
Following the methodology outlined in \cite{angeli2024acquires, keyfitz2005applied}, we introduce a perturbation to \(\mathbf{K}\) in Eq. \eqref{eq: eigen_problem}:
\[
(\mathbf{K} + \Delta \mathbf{K})(\mathbf{w} + \Delta \mathbf{w}) = (\mathcal{R}_0 + \Delta \mathcal{R}_0)(\mathbf{w} + \Delta \mathbf{w}).
\]
Expanding both sides and omitting second-order terms:
\[
\mathbf{K} \mathbf{w} + \mathbf{K} \Delta \mathbf{w} + \Delta \mathbf{K} \mathbf{w} = \mathcal{R}_0 \mathbf{w} + \mathcal{R}_0 \Delta \mathbf{w} + \Delta \mathcal{R}_0 \mathbf{w}.
\]
Subtracting $\mathbf{K} \mathbf{w} = \mathcal{R}_0 \mathbf{w}$ from both sides yields:
\[
\mathbf{K} \Delta \mathbf{w} + \Delta \mathbf{K} \mathbf{w} = \mathcal{R}_0 \Delta \mathbf{w} + \Delta \mathcal{R}_0 \mathbf{w}.
\]
Rearranging terms:
\[
\Delta \mathbf{K} \mathbf{w} = \Delta \mathcal{R}_0 \mathbf{w} + (\mathcal{R}_0 \mathbf{I} - \mathbf{K}) \Delta \mathbf{w}.
\]
To isolate $\Delta \mathcal{R}_0$, we project both sides onto the left eigenvector $\mathbf{v}^\mathrm{T}$:
\[
\mathbf{v}^\mathrm{T} \Delta \mathbf{K} \mathbf{w} = \Delta \mathcal{R}_0 \cdot \mathbf{v}^\mathrm{T} \mathbf{w} + \mathbf{v}^\mathrm{T} (\mathcal{R}_0 \mathbf{I} - \mathbf{K}) \Delta \mathbf{w}.
\]
Using the identity $\mathbf{v}^\mathrm{T} \mathbf{K} = \mathcal{R}_0 \mathbf{v}^\mathrm{T}$, we find that:
\[
\mathbf{v}^\mathrm{T} (\mathcal{R}_0 \mathbf{I} - \mathbf{K}) = \mathbf{0},
\]
which implies the last term vanishes. Thus:
\[
\Delta \mathcal{R}_0 = \frac{\mathbf{v}^\mathrm{T} \Delta \mathbf{K} \mathbf{w}}{\mathbf{v}^\mathrm{T} \mathbf{w}}.
\]
Now, consider a perturbation that affects only one entry of $\mathbf{K}$ denoted by \(k_{ij}\), then:
\[
\Delta \mathcal{R}_0 = \frac{v_i w_j}{\mathbf{v}^\mathrm{T} \mathbf{w}} \Delta k_{ij}.
\]
Taking the limit as $\Delta k_{ij}\!\to\!0$ yields
\begin{equation}
\frac{\partial \mathcal{R}_0}{\partial k_{ij}} \;=\; \frac{v_i w_j}{\mathbf{v}^\mathrm{T}\mathbf{w}},
\label{eq:sens}
\end{equation}
the sensitivity of $\mathcal{R}_0$ to the NGM entry $k_{ij}$. 
Following \cite{angeli2024acquires}, we obtain the corresponding proportional sensitivity,
\begin{equation}
S_{ij} = \frac{k_{ij}}{\mathcal{R}_0}
\frac{\partial \mathcal{R}_0}{\partial k_{ij}}
    \label{eq:elasticity}
\end{equation}
In our framework, the entries of $\mathbf{K}$ depend on contact data and epidemiological rates, so we treat contact entries as \emph{low-level parameters} and obtain derivatives via the chain rule (Eq.(~\ref{eq: r0_cm})). Consequently, contact-level derivatives do not admit a closed form like \eqref{eq:sens} and are computed through the chain rule.

\section{Automatic Differentiation to Calculate Gradients}

In our framework, a \textit{computational graph} is the sequence of mathematical operations that maps the independent entries of the population-scaled contact matrix \( C^{\mathrm{pop}} \) to the epidemiological outcomes of interest, such as the basic reproduction number \(\mathcal{R}_0\). Each node of the graph represents an algebraic operation (e.g., reciprocity restoration, scaling, matrix multiplication, or eigenvalue computation), and the directed edges represent the flow of intermediate results.

The inputs to this graph are the \(n_a \times n_a\) elements of the fully adjusted and population-scaled contact matrix, \( C^{\mathrm{pop}} \in \mathbb{R}^{n_a \times n_a} \), as defined in Eq.~\eqref{pop_sum}. This matrix, which encodes the total contact rates between age groups normalized by the overall population, forms the foundation of the computational graph. It serves as the starting point for automatic differentiation, enabling gradient-based sensitivity analysis of transmission dynamics and outcome measures.

In the first phase, we treat only the upper-triangular (including diagonal) entries of \(C^{\mathrm{pop}}\) as independent parameters and complete the matrix by reflecting them to the lower triangle. The resulting symmetry arises from the reciprocity-restoring averaging at the total-contacts level (Eq.~\eqref{total}), not from assuming per-capita symmetry in the raw data. Subsequently, this matrix is converted into the adjusted contact matrix \(\overline{C}\), as defined in Eq.~\eqref{sym_contact}. The matrix \(\overline{C} \in \mathbb{R}^{n_a \times n_a}\) serves as the contact input to the transmission model.

The computation graph then proceeds by feeding \(\overline{C}\) into an age-structured compartmental model to compute the next-generation matrix $\mathbf{K}$. Depending on the outcome of interest, $\mathbf{K}$ may be further transformed (e.g., by incorporating hospitalization, ICU, or mortality probabilities), and its dominant eigenvalue is extracted to compute \(\mathcal{R}_0\). 
Automatic differentiation is applied at this stage using the \texttt{PyTorch} library. We perform backpropagation through the entire graph to compute the gradient of \(\mathcal{R}_0\) with respect to the independent entries of \(C^{\mathrm{pop}}\). These gradients quantify the sensitivity of epidemic outcomes to variations in specific age-specific contact rates, forming the basis of cumulative sensitivity analyses.

\section{Epidemic models}

\subsection{Pitman et al. model}\label{sec:pitman}

The SEIR model in Eq. (\ref{eq: seir}) describes the progression of individuals through susceptible (\(S\)), exposed (\(E\)), infectious (\(I\)), and recovered (\(R\)) states, with key parameters: rate of latent infection \(\alpha\), and recovery rate \(\gamma\) as adapted from \cite{pitman2012estimating}.
The corresponding system of ordinary differential equations (ODEs) is given by:
\begin{equation}
\begin{aligned}
    S_i'(t) &= -\beta \cdot \frac{S_i(t)}{N_i} 
    \sum_{j=1}^{n} c_{i,j} \cdot I_j(t), \\[6pt]
    E_i'(t) &= \beta \cdot \frac{S_i(t)}{N_i} 
    \sum_{j=1}^{n} c_{i,j} \cdot I_j(t) - \alpha E_i(t), \\[6pt]
    I_i'(t) &= \alpha E_i(t) - \gamma I_i(t), \\[6pt]
    R_i'(t) &= \gamma I_i(t).
\end{aligned}
\label{eq: seir}
\end{equation}
The model incorporates age-specific contact patterns through the matrix $c_{i,j}
$, where contacts are based on pre-pandemic estimates. The parameter sets used for both the full-resolution and aggregated models are summarized in Table~\ref{tab: seir}.

\begin{table}[!h]
\small
    \centering
    \begin{tabular}{l c c c}
        \toprule
        Parameter & Value & Description \\
        \midrule
        $\alpha$ & 0.5 & Rate of latent becoming infectious \\
        $\gamma$ & 0.5 & Recovery rate \\
        $\overline{\mathcal{R}}_0$ & 1.8 & Baseline Reproduction number \\
        \bottomrule
    \end{tabular}
    \caption{\small
    Model parameters are derived from \cite{pitman2012estimating}, using UK population and contact data from \cite{mossong2008} to compute $\mathbf{K}$ and $\mathcal{R}_0$ based on Eq. (\ref{eq: seir}).}   
    \label{tab: seir}
\end{table}

\subsection{R{\"o}st et al. model}\label{sec:rost}

\begin{figure}[H]
    \centering
    \includegraphics[width=\textwidth]{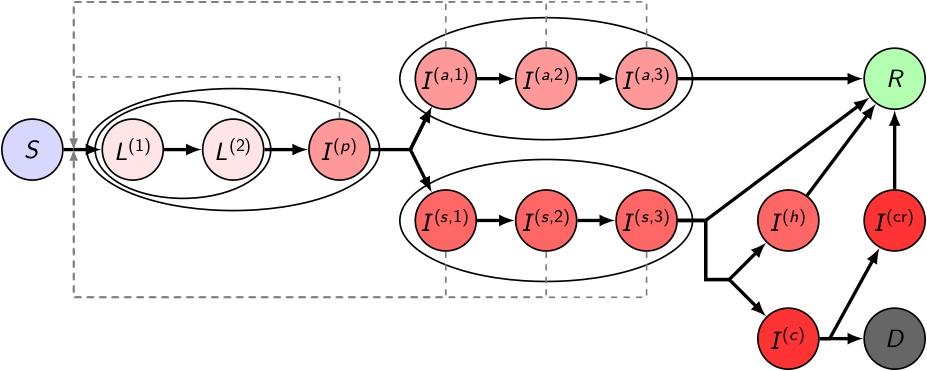}
    \caption{\small Compartmental transmission diagram of the age-structured COVID-19 model, adapted from \cite{Rost}. Solid black arrows represent transitions between epidemiological compartments, while dashed gray arrows indicate potential infection pathways leading into the infectious stages. The model includes age-dependent branching probabilities for asymptomatic progression ($p_i$), hospitalization ($h_i$), critical care requirement ($\xi_i$), and fatality ($\mu_i$).
}  
    \label{fig:diag}
\end{figure}

In this COVID-19 model, the population is divided into multiple compartments, each representing a distinct stage of infection. These compartments include susceptible individuals (\(S\)), who are at risk of contracting the virus, and latent individuals (\(L\)), who have been infected but are not yet symptomatic. As the infection progresses, individuals transition into the presymptomatic stage (\(I^{(p)}\)) before the onset of noticeable symptoms. Some individuals may remain asymptomatic (\(I^{(a)}\)), carrying the virus without exhibiting symptoms. Others will develop symptoms of varying severity, classifying them as symptomatic individuals (\(I^{(s)}\)). Those who experience severe symptoms and require hospitalization are categorized as hospitalized individuals (\(I^{(h)}\)). Among these, the most critically ill patients who require intensive care are placed in the intensive care unit (ICU) compartment (\(I^{(c)}\)). Finally, individuals who recover from the infection and gain immunity are classified as recovered (\(R\)), while those who succumb to the disease are placed in the deceased (\(D\)) compartment. The compartments and their transitions are depicted in Fig. \ref{fig:diag}, and the complete set of model equations is provided in (\ref{eq: rost}). 

The following set of ordinary differential equations governs the dynamics of the system:
\begin{equation}
\begin{aligned}
S_i'(t) &= -\beta \cdot \frac{S_i(t)}{N_i} \cdot   \\
&\quad
\sigma_i 
\sum_{j=1}^{n} c_{i,j} 
\left[ I^{(p)}_j(t) + \mathrm{inf}^{(a)} \sum_{m=1}^3 I^{(a,m)}_j(t) + \sum_{m=1}^3 I^{(s,m)}_j(t) \right], \\[6pt]
{L^{(1)}}_i'(t) &= \beta \cdot \frac{S_i(t)}{N_i} \cdot  \\
&\quad
\sigma_i 
\sum_{j=1}^{n} c_{i,j} 
\left[ I^{(p)}_j(t) + \mathrm{inf}^{(a)} \sum_{m=1}^3 I^{(a,m)}_j(t) + \sum_{m=1}^3 I^{(s,m)}_j(t) \right] 
- 2\alpha^{(L)} L^{(1)}_i(t), \\[6pt]
{L^{(2)}}_i'(t) &= 2\alpha^{(L)} L^{(1)}_i(t) - 2\alpha^{(L)} L^{(2)}_i(t), \\[6pt]
{I^{(p)}}_i'(t) &= 2\alpha^{(L)} L^{(2)}_i(t) - \alpha^{(p)} I^{(p)}_i(t), \\[6pt]
{I^{(a,1)}}_i'(t) &= p_i \alpha^{(p)} I^{(p)}_i(t) - 3\gamma^{(a)} I^{(a,1)}_i(t), \\
{I^{(a,2)}}_i'(t) &= 3\gamma^{(a)} I^{(a,1)}_i(t) - 3\gamma^{(a)} I^{(a,2)}_i(t), \\
{I^{(a,3)}}_i'(t) &= 3\gamma^{(a)} I^{(a,2)}_i(t) - 3\gamma^{(a)} I^{(a,3)}_i(t), \\[6pt]
{I^{(s,1)}}_i'(t) &= (1 - p_i) \alpha^{(p)} I^{(p)}_i(t) - 3\gamma^{(s)} I^{(s,1)}_i(t), \\
{I^{(s,2)}}_i'(t) &= 3\gamma^{(s)} I^{(s,1)}_i(t) - 3\gamma^{(s)} I^{(s,2)}_i(t), \\
{I^{(s,3)}}_i'(t) &= 3\gamma^{(s)} I^{(s,2)}_i(t) - 3\gamma^{(s)} I^{(s,3)}_i(t), \\[6pt]
{I^{(h)}}_i'(t) &= h_i (1 - \xi_i) \cdot 3\gamma^{(s)} I^{(s,3)}_i(t) - \gamma^{(h)} I^{(h)}_i(t), \\[6pt]
{I^{(c)}}_i'(t) &= h_i \xi_i \cdot 3\gamma^{(s)} I^{(s,3)}_i(t) - \gamma^{(c)} I^{(c)}_i(t), \\
{I^{(\mathrm{cr})}}_i'(t) &= (1 - \mu_i) \gamma^{(c)} I^{(c)}_i(t) - \gamma^{(\mathrm{cr})} I^{(\mathrm{cr})}_i(t), \\[6pt]
R_i'(t) &= 3\gamma^{(a)} I^{(a,3)}_i(t) + (1 - h_i) 3\gamma^{(s)} I^{(s,3)}_i(t) 
+ \gamma^{(h)} I^{(h)}_i(t) + \gamma^{(\mathrm{cr})} I^{(\mathrm{cr})}_i(t), \\[6pt]
D_i'(t) &= \mu_i \gamma^{(c)} I^{(c)}_i(t).
\label{eq: rost}
\end{aligned}
\end{equation}
In this model, individuals are divided into age groups indexed by \( i \in \{1, \ldots, n\} \), where \( \beta \) denotes the probability of transmission upon contact. Each parameter indexed by \( i \) corresponds to a specific age group. The model leverages the contact matrix shown in Fig.~\ref{fig:contacts} to compute sensitivity values relevant to transmission dynamics.  
A complete list of model parameters used for both the full-resolution and aggregated versions of the model is provided in the accompanying Zenodo repository \cite{code}.




\end{appendices}



\begin{thebibliography}{99}

\bibitem{abrams}
Abrams, S., Wambua, J., Santermans, E., Willem, L., Kuylen, E., Coletti, P., ... \& Hens, N. (2021). Modelling the early phase of the Belgian COVID-19 epidemic using a stochastic compartmental model and studying its implied future trajectories. Epidemics, 35, 100449.


\bibitem{adu2022quantifying}
Adu, P., Binka, M., Mahmood, B., Jeong, D., Buller-Taylor, T., Damascene, MJ, ... \& Janjua, N. (2022). Quantifying contact patterns: development and characteristics of the British Columbia COVID-19 population mixing patterns survey. International Journal of Infectious Diseases , 116 , S30-S31.


\bibitem{anderson}
Anderson, R. M., \& May, R. M. (1985). Age-related changes in the rate of disease transmission: implications for the design of vaccination programmes. Epidemiology \& Infection, 94(3), 365-436.


\bibitem{angeli2024acquires}
Angeli, L., Caetano, CP, Franco, N., Abrams, S., Coletti, P., Van Nieuwenhuyse, I., ... \& Hens, N. (2024). Who acquires infection from whom? A sensitivity analysis of transmission dynamics during the early phase of the COVID-19 pandemic in Belgium. Journal of Theoretical Biology , 581 , 111721.

\bibitem{bengio}
Bengio, Y., Ippolito, D., Janda, R., Jarvie, M., Prud'homme, B., Rousseau, JF, ... \& Yu, YW (2021). Inherent privacy limitations of decentralized contact tracing apps. Journal of the American Medical Informatics Association , 28 (1), 193-195.


\bibitem{bokanyi}
Bokányi, E., Vizi, Z., Koltai, J., Röst, G., \& Karsai, M. (2023). Real-time estimation of the effective reproduction number of COVID-19 from behavioral data. Scientific Reports, 13(1), 21452.


\bibitem{cao}
Cao, Y., Li, S., \& Petzold, L. (2002). Adjoint sensitivity analysis for differential-algebraic equations: algorithms and software. Journal of computational and applied mathematics , 149 (1), 171-191.

\bibitem{diekmann}
Diekmann, O., Heesterbeek, J. A. P., \& Roberts, M. G. (2010). The construction of next-generation matrices for compartmental epidemic models. Journal of the royal society interface, 7(47), 873-885.

\bibitem{evans}
Korir, E. K., \& Vizi, Z. (2022, November). Clustering of countries based on the associated social contact patterns in epidemiological modelling. In International Symposium on Mathematical and Computational Biology (pp. 253-271). Cham: Springer Nature Switzerland.

\bibitem{fumanelli}
Fumanelli, L., Ajelli, M., Manfredi, P., Vespignani, A., \& Merler, S. (2012). Inferring the structure of social contacts from demographic data in the analysis of infectious diseases spread.

\bibitem{knipl}
Knipl, D. H., \& Röst, G. (2009). Influenza models with Wolfram Mathematica. Interesting Mathematical Problems in Sciences and Everyday Life, 1-24.


\bibitem{koltai}
Koltai, J., Vásárhelyi, O., Röst, G., \& Karsai, M. (2022). Reconstructing social mixing patterns via weighted contact matrices from online and representative surveys. Scientific reports, 12(1), 4690.

\bibitem{korir}
Korir, E. K., \& Vizi, Z. (2024). Clusters of African countries based on the social contacts and associated socioeconomic indicators relevant to the spread of the epidemic. Journal of Mathematics in Industry, 14(1), 24.

\bibitem{mossong2008}
Mossong, J., Hens, N., Jit, M., Beutels, P., Auranen, K., Mikolajczyk, R., ... \& Edmunds, W. J. (2008). Social contacts and mixing patterns relevant to the spread of infectious diseases. PLoS medicine, 5(3), e74.

\bibitem{munday}
Munday, J. D., Abbott, S., Meakin, S., \& Funk, S. (2023). Evaluating the use of social contact data to produce age-specific short-term forecasts of SARS-CoV-2 incidence in England. PLoS Computational Biology, 19(9), e1011453.


\bibitem{prem}
Prem, K., Zandvoort, KV, Klepac, P., Eggo, RM, Davies, NG, Center for the Mathematical Modeling of Infectious Diseases COVID-19 Working Group, ... \& Jit, M. (2021). Projecting contact matrices in 177 geographical regions: an update and comparison with empirical data for the COVID-19 era. PLoS computational biology , 17 (7), e1009098.


\bibitem{Rost}
Röst, G., Bartha, F. A., Bogya, N., Boldog, P., Dénes, A., Ferenci, T., ... \& Oroszi, B. (2020). Early phase of the COVID-19 outbreak in Hungary and post-lockdown scenarios. Viruses, 12(7), 708.

\bibitem{sorensen}
Sorensen, RJD, Barber, RM, Pigott, DM, Carter, A., Spencer, CN, Ostroff, SM, ... \& Murray, C. (2022). Variation in the COVID-19 infection-fatality ratio by age, time, and geography during the pre-vaccine era: A systematic analysis. The Lancet , 399 (10334), 1469-1488.


\bibitem{wallinga}
Wallinga, J., Teunis, P., \& Kretzschmar, M. (2006). Using data on social contacts to estimate age-specific transmission parameters for respiratory-spread infectious agents. American journal of epidemiology , 164 (10), 936-944.

\bibitem{le2018characteristics}
Le Polain de Waroux, O., Cohuet, S., Ndazima, D., Kucharski, A. J., Juan-Giner, A., Flasche, S., ... \& Edmunds, W. J. (2018). Characteristics of human encounters and social mixing patterns relevant to infectious diseases spread by close contact: a survey in Southwest Uganda. BMC infectious diseases, 18, 1-12.

\bibitem{kiti2014quantifying}
Kiti, M. C., Kinyanjui, T. M., Koech, D. C., Munywoki, P. K., Medley, G. F., \& Nokes, D. J. (2014). Quantifying age-related rates of social contact using diaries in a rural coastal population of Kenya. PloS one, 9(8), e104786.

\bibitem{ajelli2017estimating}
Ajelli, M., \& Litvinova, M. (2017). Estimating contact patterns relevant to the spread of infectious diseases in Russia. Journal of theoretical biology, 419, 1-7.


\bibitem{melegaro2017social}
Melegaro, A., Del Fava, E., Poletti, P., Merler, S., Nyamukapa, C., Williams, J., ... \& Manfredi, P. (2017). Social contact structures and time use patterns in the Manicaland Province of Zimbabwe. PloS one, 12(1), e0170459.



\bibitem{kumar2018interacts}
Kumar, S., Gosain, M., Sharma, H., Swetts, E., Amarchand, R., Kumar, R., ... \& Krishnan, A. (2018). Who interacts with whom? Social mixing insights from a rural population in India. Plos one , 13 (12), e0209039.


\bibitem{read2014social}
Read, J. M., Lessler, J., Riley, S., Wang, S., Tan, L. J., Kwok, K. O., ... \& Cummings, D. A. (2014). Social mixing patterns in rural and urban areas of southern China. Proceedings of the Royal Society B: Biological Sciences, 281(1785), 20140268.


\bibitem{horby2011social}
Horby, P., Thai, PQ, Hens, N., Yen, NTT, Mai, LQ, Thoang, DD, ... \& Hien, NT (2011). Social contact patterns in Vietnam and implications for the control of infectious diseases. PloS one , 6 (2), e16965.

\bibitem{grijalva2015household}
Grijalva, CG, Goeyvaerts, N., Verastegui, H., Edwards, KM, Gil, AI, Lanata, CF, ... \& RESPIRA PERU project. (2015). A household-based study of contact networks relevant for the spread of infectious diseases in the highlands of Peru. PloS one , 10 (3), e0118457.

\bibitem{pitman2012estimating}
Pitman, R. J., White, L. J., \& Sculpher, M. (2012). Estimating the clinical impact of introducing paediatric influenza vaccination in England and Wales. Vaccine, 30(6), 1208-1224.

\bibitem{code}
Code and data repository for Eigenvector-Based Sensitivity Analysis. Available at: \url{https://zenodo.org/records/17288964}.

\bibitem{hamby1994review}
Hamby, D. M. (1994). A review of techniques for parameter sensitivity analysis of environmental models. Environmental monitoring and assessment, 32, 135-154.

\bibitem{sobol1990sensitivity}
Sobol', IYM (1990). On sensitivity estimation for nonlinear mathematical models. Matematicheskoe modelirovanie , 2 (1), 112-118.


\bibitem{morris1992factorial}
Morris, M. D. (1991). Factorial sampling plans for preliminary computational experiments. Technometrics, 33(2), 161-174.

\bibitem{helton2000sampling}
Helton, JC, \& Davis, FJ (2000). Sampling-based methods for uncertainty and sensitivity analysis (No. SAND99-2240). Sandia National Lab. (SNL-NM), Albuquerque, NM (United States); Sandia National Lab. (SNL-CA), Livermore, CA (United States).

\bibitem{sensitivity}
Vizi, Z., Korir, EK, Bogya, N., Rosztóczy, C., Makay, G., \& Boldog, P. (2025). Age Group Sensitivity Analysis of Epidemic Models: Investigating the Impact of Contact Matrix Structure. arXiv preprint arXiv:2502.19206 .

\bibitem{heder2022past}
Héder, M., Rigó, E., Medgyesi, D., Lovas, R., Tenczer, S., Farkas, A., ... \& Kacsuk, P. (2022). The past, present and future of the ELKH cloud. INFORMÁCIÓS TÁRSADALOM: TÁRSADALOMTUDOMÁNYI FOLYÓIRAT, 22(2), 128-137.


\bibitem{keyfitz2005applied}
Keyfitz, N., \& Caswell, H. (2005). Applied mathematical demography (Vol. 47). New York: Springer.


\end{thebibliography}

\end{document}